\documentclass[english,aps,prb,twocolumn,footinbib,longbibliography,superscriptaddress,floatfix]{revtex4-1}

\usepackage[english]{babel}
\usepackage{amssymb,amsmath}
\usepackage[pdftex]{graphicx}
\usepackage{color}
\usepackage[usenames,dvipsnames]{xcolor}
\usepackage{hyperref} 
\hypersetup{%
	linktoc=page,%
	colorlinks=true,%
	linkcolor=RoyalBlue,%
	citecolor=OliveGreen,%
	urlcolor=OliveGreen,%
	linkbordercolor=red,%
	pdfborderstyle={/S/U/W 1},%
}
\usepackage{float}
\usepackage{setspace}

\begin{document}

\title{Arrow of Time and its Reversal on IBM Quantum Computer}

\author{G. B. Lesovik}
\affiliation{Moscow Institute of Physics and Technology, Institutskii per. 9, Dolgoprudny, 141700, Moscow District, Russia}

\author{I. A. Sadovskyy}
\affiliation{Materials Science Division, Argonne National Laboratory, 9700 S. Cass Av., Argonne, IL 60637, USA}
\affiliation{Computation Institute, University of Chicago, 5735 S. Ellis Av., Chicago, IL 60637, USA}

\author{M. V. Suslov}
\affiliation{Moscow Institute of Physics and Technology, Institutskii per. 9, Dolgoprudny, 141700, Moscow District, Russia}

\author{A. V. Lebedev}
\affiliation{Theoretische Physik, ETH Z\"urich, Wolfgang-Pauli-Strasse 27, CH-8093 Z\"urich, Switzerland}

\author{V. M. Vinokur}
\affiliation{Materials Science Division, Argonne National Laboratory, 9700 S. Cass Av., Argonne, IL 60637, USA}

\begin{abstract}
Uncovering the origin of the arrow of time remains a fundamental scientific challenge. Within the framework of statistical physics, this problem was inextricably associated with the Second Law of Thermodynamics, which declares that entropy growth proceeds from the system's entanglement with the environment. It remains to be seen, however, whether the irreversibility of time is a fundamental law of nature or whether, on the contrary, it might be circumvented. Here we show that, while in nature the complex conjugation needed for time reversal is exponentially improbable, one can design a quantum algorithm that includes complex conjugation and thus reverses a given quantum state. Using this algorithm on an IBM quantum computer enables us to experimentally demonstrate a backward time dynamics for an electron scattered on a two-level impurity.
\end{abstract}

\maketitle

\begin{spacing}{0}
\tableofcontents
\end{spacing}

\section{Introduction}

For decades, researchers have sought to understand how the irreversibility of the surrounding world emerges from the seemingly time-symmetric, fundamental laws of physics.\cite{Lloyd:2013} This question is as old as classical statistical mechanics,\cite{Kelvin:1857, Maxwell:1860, Boltzmann:1872, Boltzmann:1896, Lebowitz:1999} which itself represented an attempt to solve this enigma. Quantum mechanics stepped into the breach with two important conjectures. The first, independently proposed by both Landau\cite{Landau:1927} and von Neumann,\cite{Neumann:1929} postulated that the process of macroscopic measurement creates irreversibility. The second, due to Wigner,\cite{Wigner:1932} posited that time reversal operation is anti-unitary because it requires complex conjugation. More recently, the ontological status of time-reversal symmetry of quantum mechanics as a version of probabilistic theory was discussed in detail in Ref.~\onlinecite{Holster:2003, Oreshkov:2015}. Because of the requirement for complex conjugation, the universal time reversal operation lies outside the quantum realm and does not spontaneously appear in nature. The above conjectures, though crucial in their own right, nonetheless represent two different keys to the same lock. In this paper, we uncover the interrelationships between these seemingly disparate keys. It is known that dissipation is a particular case of unitary evolution accompanied by entanglement. Entanglement, in turn, complicates quantum states by involving more and more degrees of freedom~--- an increase in complexity that renders spontaneous time reversal highly improbable. Our perspective refines the Landau-von Neumann insight into irreversibility. As it turns out, the measurement process can be described as the joint unitary evolution of the quantum system and the macroscopic measuring device. The resulting macroscopic entanglement gives rise to the insurmountable complexity of the reversal procedure. We further show that, unexpectedly, even the evolution of single- or two-particle states in free space generates complexity that renders spontaneous time reversal either highly improbable or actually impossible. This expresses the fact that the Schr\"{o}dinger equation determining evolution of quantum systems implicitly entails irreversibility~--- hence the arrow of time in nature.

\begin{figure*}
\centering
\includegraphics[width=0.9\linewidth]{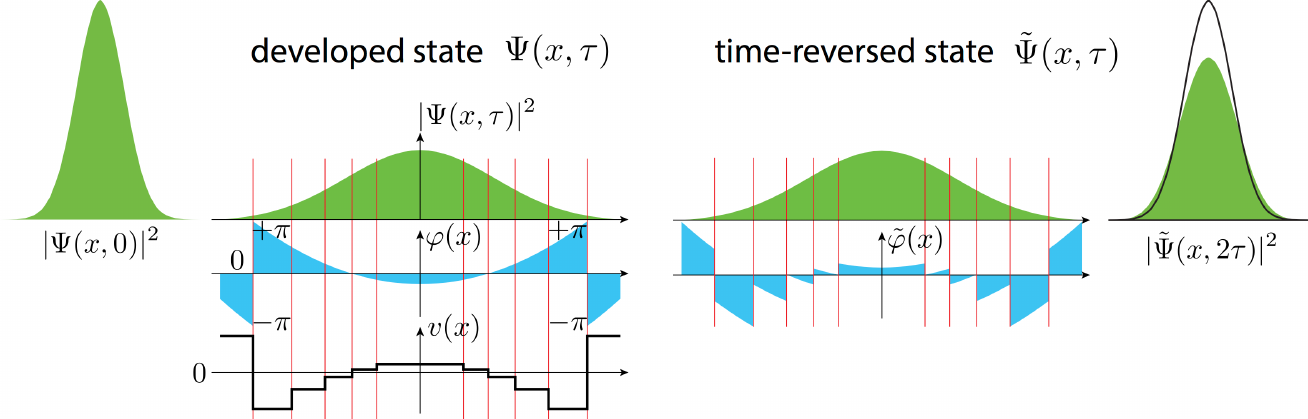}
\caption{Time reversal procedure for a Gaussian wave-packet $\Psi(x,0) \propto e^{-x^2/2\sigma^2}$, $\sigma = 1$ (a.u.). The wave-packet spreads $\Psi(x,0) \to \Psi(x,\tau)$ according to a quadratic Hamiltonian $\hat{p}^2/2m$ during the time interval $\tau=3m\sigma^2/\hbar$. At the moment $\tau$ the system is exposed to the fast step-wise electromagnetic potential fluctuation $v(x)$ (second panel). The fluctuation approximately (with the precision corresponding to the density of partitioning points) conjugates the phase of the wave-function: $\varphi(x,\tau^{-0}) \to \tilde\varphi(x,\tau^{+0}) = \varphi(x,\tau^{-0}) + ev(x,\tau)\delta\tau/\hbar$ (third panel). The prepared time-reversed state $\tilde\Psi(x,\tau)$ then freely evolves during the same time interval $\tau$ and arrives to the squeezed state $\tilde\Psi(x,2\tau)$ (fourth panel). The resulting state $\tilde\Psi(x,2\tau)$ has $86\%$ overlap with the initial state $\Psi(x,0)$ shown as an empty envelope curve in the fourth panel.
}
\label{fig:wave_packet_reversal}
\end{figure*}

\section{Reversal of the spreading wave packet}

That in quantum mechanics in order to execute a time reversal operation one has to perform complex conjugation of the wave function, implies that the time reversal operator $\mathcal{\hat T}$ is a product of a complex conjugation operator $\mathcal{\hat K}$ and a unitary rotation $\hat{U}_R$, i.e. $\mathcal{\hat T} = \hat{U}_R \, \mathcal{\hat K}$, where for any $\Psi$, $\mathcal{\hat K}\Psi = \Psi^*$. This operation not only reflects velocities like in the classical physics, but also reverses phases of the wave function components. A general universal operation that can reverse any arbitrary wave function, does not exist in nature. Yet, some special $\Psi$-dependent operation such that $\hat{U}_{\Psi}\Psi=\Psi^*$ can exist and below we explicitly construct such an operation for a system of qubits. To that end, one has to design a supersystem that is external with respect to the system of interest and which is capable to implement the purposeful manipulating on the given system. In nature, in the simplest case of a single particle, the role of such a supersystem can be taken up, for example, by the fluctuating electromagnetic field. To gain an insight into how this works, let us consider a wave packet corresponding to the particle with the square energy dispersion, $\varepsilon=p^2/2m$, where $p$ is the particle momentum and $m$ is the particle mass, propagating in space, see Fig.~\ref{fig:wave_packet_reversal}. The electromagnetic field is assumed to be predominantly weak except for rare fluctuations. Thus, the spreading of the wave packet is coherent. At large times $\tau$ the wave packet spreads as
\begin{equation}
	\Psi(x,\tau) 
	\simeq \frac{f(xm/\hbar\tau)}{\sqrt{2\pi\hbar\tau/m}} \,
	e^{i m x^2/2\hbar\tau},
	\label{eq:wave_packet}
\end{equation}
where $f(q)$ Fourier image of the initial spatial wave function. The phase of $\Psi$ changes as a result of the action of the fast fluctuation of an external potential, i.e. with the potential that changes on the times much shorter than the characteristic time of the phase change. To set the fluctuation that complex conjugates $\Psi$, let us divide the coordinate space into a large number of the elemental cells $\delta x_n$ where a wave function's phase $\varphi(x,\tau)$ changes slowly and look for a fast electromagnetic potential fluctuation $V(x,t)$ which is smooth on the cell's scale and reverts the phase of the wave function: $\int dt\, eV(x_n,t)/\hbar = -2\varphi(x_n,\tau)$. If during the $\tau$ the wave packet~\eqref{eq:wave_packet} has spread from the size $L_0$ to the size $L_\tau = \hbar\tau/m L_0$, it would require $N \sim \epsilon^{-1/2} (L_\tau/L_0)$ elementary cells to approximately revert the quantum state $\Psi(x,\tau) \to \tilde\Psi^*(x,\tau)$ with the probability $1-\epsilon$: $|\langle \tilde\Psi^*(x,\tau)|\Psi^*(x,\tau) \rangle|^2 = 1 - \epsilon$, see Appendix~\ref{sec:wave_packet_reversal}. Then the probability of the spontaneous reversal, i.e. the probability of the appearance of the required electromagnetic potential fluctuation, estimates as $2^{-N}$. Now we determine the typical time scale $\tau$ on which the spontaneous time reversal of a wave-packet can still occur within the universe lifetime $t_U \sim 4.3 \times 10^{17}$\,sec. The latter is obtained from the estimate $2^{-N}\simeq \tau/t_U$, where the number of cells $N \sim \epsilon^{-1/2}\, (\langle E\rangle\tau/\hbar)$ is expressed through the average particle energy $\langle E\rangle = \hbar^2/mL_0^2$. As a typical average energy of the wave-packet we take the energy corresponding to the current universe temperature $2.72$\,K and arrive at $\tau \simeq 6 \times 10^{-11}$\,sec. One thus sees that even in the discussed simplest possible example of a single quantum particle the time reversal is already a daunting task where even with the GHz rate of attempts, the required fluctuation is not observable within the universe lifetime.

Now we consider a more complex example and demonstrate that a separable state $\Psi(x_1,x_2) = |\psi_1(x_1)\psi_2(x_2)| \, e^{i\varphi_1(x_1) + i\varphi_2(x_2)}$ of two particles can not be reverted by classical field fluctuations in the case where particle's wave functions overlap. Let all particles have the same electric charge $q$ and interact with a classical electric potential $v(x,t)$. The potential fluctuations produce phase shifts $\int dt \, q v(x,t)/\hbar$. Accordingly the proper fluctuations capable to reverse the quantum state should satisfy the condition $\varphi_1(x_1) + \varphi_2(x_2) + \int dt \, [q v(x_1,t) + q v(x_2,t)] / \hbar = - \varphi_1(x_1) - \varphi_2(x_2)$. For $x_1=x_2$ it implies $ \int dt \, q v(x,t) / \hbar = -\varphi_1(x)-\varphi_2(x)$, and therefore at $x_1 \neq x_2$ one has to satisfy the condition $\varphi_2(x_2) + \varphi_1(x_1) = \varphi_2(x_1) +\varphi_1(x_2)$ which, in general, does not hold.

Quantum entanglement introduces the next level of complexity for the time-reversal procedure. Consider a two-particle state
$\Psi(x_1,x_2)=|\Psi(x_1,x_2)| e^{i\varphi(x_1,x_2)}$ with the non-separable phase function $\varphi(x_1,x_2) = a_1(x_1)b_1(x_2) + a_2(x_1) b_2(x_2)$. In this situation even for the non-overlapping particles with $\Psi(x_1,x_2) = 0$ for $x_1=x_2$ the two-particle state can not be reversed by an interaction with classical fields. Let one access the particles by different fields which induce separate phase shifts $\Psi(x_1,x_2) \to \Psi(x_1,x_2) e^{i\phi_1(x_1) + i\phi_2(x_2)}$. The induced phase shifts should satisfy the relation: $\phi_1(x_1) + \phi_2(x_2) = -2\varphi(x_1,x_2)$, therefore for any three points $x_1 \neq x_2 \neq x_3$ the following conditions should hold
\begin{align*}
	& \phi_1(x_1) + \phi_2(x_2) = -2 \bigl[ a_1(x_1)b_1(x_2) + a_2(x_1)b_2(x_2) \bigr], \\
	& \phi_1(x_1) + \phi_2(x_3) = -2 \bigl[ a_1(x_1)b_1(x_3) + a_2(x_1)b_2(x_3) \bigr].
\end{align*}
Subtracting these relations one gets $\phi_2(x_2) -\phi_2(x_3) = -2a_1(x_1) [b_1(x_2)-b_1(x_3)] -2a_2(x_1) [b_2(x_2) - b_2(x_3)]$ where the left hand side does not depend on $x_1$ and therefore one has to assume $a_1$ and $a_2$ to be constant. This, however, contradicts the non-separability assumption for $\varphi(x_1,x_2)$.

An entangled two-particle state with a non-separable phase function can naturally emerge as a result of scattering of two localized wave-packets.\cite{Lebedev:2008} However, as we have seen, the generation of the time-reversed state, where a particle gets disentangled in the course of its forward time evolution, requires specific two-particle operations which, in general, cannot be reduced to a simple two-particle scattering.

From the above consideration we can draw important conclusions about the origin of the arrow of time:
\textit{
\begin{enumerate}
\item[(i)] For the time reversal one needs a supersystem manipulating the system in question. In the most of the cases, such a supersystem cannot spontaneously emerge in nature.
\item[(ii)] Even if such a supersystem would emerge for some specific situation, the corresponding spontaneous time reversal typically requires times exceeding the universe lifetime.
\end{enumerate}
}

A matter-of-course supersystem of that kind is implemented by the so-called universal quantum computer. It is capable to efficiently simulate unitary dynamics of any physical system endowed with local interactions.\cite{Lloyd:1996} A system's state is encoded into the quantum state of the computer's qubit register and its evolution is governed by the quantum program, a sequence of the universal quantum gates applied to the qubit register. In what follows, we first formulate general principles of constructing time-reversal algorithms on quantum computers and, in the next section, present a practical implementation of a few-qubit algorithm that enabled experimental time reversal procedure on the IBM quantum computer.

\section{General time reversal algorithms} \label{sec:reversal_algorithms}

Consider a quantum system initially prepared in the state $\Psi(t=0)$ and let it evolve during the time $\tau$ into the state $\Psi(\tau) = e^{-i\mathcal{\hat H}\tau/\hbar} \Psi(0)$. Let us find a minimal size of a qubit register needed to simulate the dynamics of a system $\Psi(0) \to \Psi(\tau)$ with a given fidelity $1-\epsilon$. Let us choose a finite set of time instances $t_i \in[0,\tau]$, $i=0,\ldots \mathcal{N}^\prime$ subject to a condition $|\langle \Psi(t_i)| \Psi(t_{i+1}\rangle)|^2 = 1-\epsilon$ with $t_0 = 0$ for some small $\epsilon >0$. Then at any time instant $t\in [0,\tau]$ a state $\Psi(t)$ can be approximated by the discrete set of states $\{ \Psi(t_i), i = 0, \ldots, \mathcal{N}^\prime\}$ with the fidelity $1-\epsilon$. The set of states $\{ \Psi(t_i)\}$ spans the Hilbert subspace $\mathcal{S} $ of the dimension $\mathcal{N} \leq \mathcal{N}^\prime$. Therefore, $\mathcal{N}$ basis vectors $|e_i\rangle \in \mathcal{S}$ can be represented by $\mathcal{N}$ orthogonal states of the qubit register, $|e_i\rangle \to |\vec{b}_i\rangle \equiv |b_0b_1\ldots \rangle$. The corresponding qubit Hamiltonian $\hat{H}$ which mimics the original Hamiltonian $\mathcal{\hat H}$ is then defined by the relation $(\hat{H})_{ij} \equiv \langle \vec{b}_i| \hat{H}| \vec{b}_j\rangle = \langle e_i| \mathcal{\hat H}|e_j\rangle$.

Below we introduce two encoding procedures $|e_i\rangle \to |\vec{b}_i\rangle$. In the first, \textit{sparse} coding approach, one assigns a separate qubit to each state $|e_i\rangle$, $i \in[0,\mathcal{N}-1]$ and encodes the state $\psi(\tau)$ into the $\mathcal{N}$-qubit state $|\psi\rangle = \sum_{i=0}^{\mathcal{N}-1} \psi_i\, |0_0\ldots 1_i \ldots 0_{\mathcal{N}-1}\rangle$. The second approach is a \textit{dense} coding scheme where one records the state $\psi(\tau)$ into a state of $n = \mbox{int}[\log_2(\mathcal{N})]\!+\!1$ qubits $|\psi\rangle = \sum_{i=0}^{\mathcal{N}-1} \psi_i\, |i\rangle$, where $|i\rangle \equiv |b_0 \ldots b_{n-1}\rangle$ is a computational basis state corresponding a binary representation of the number $i = \sum_{k=0}^{n-1} b_k 2^{n-1-k}$.

A time-reversal operation $\hat{R}$ of the qubit register can be presented as a product $\hat{R} = \hat{U}_R \hat{K}$ of the complex conjugation operator $\hat{K}$, $\hat{K}(\psi_i|i\rangle) \equiv \psi_i^*|i\rangle$, and some unitary operator $\hat{U}_R$, whose form is defined by the Hamiltonian $\hat{H}$, $\hat{U}_R = \hat{U}_H^\dagger \hat{U}_H^*$, where $\hat{H} = \hat{U}_H^\dagger \mbox{diag}\{ E_1 \ldots E_n\} \hat{U}_H$, see Appendix~\ref{sec:qubit_reversal}. Therefore, in order to implement the time-reversal operation $\hat{R}$ one needs to know the Hamiltonian $\hat{H}$ explicitly. Note, that quantum computer is able to simulate unitary dynamics governed by an arbitrary Hamiltonian including those that do not correspond any physical system (for example, some non-local Hamiltonian). It is known, that the joint transformation of the charge conjugation, parity inversion, and time reversal is considered as an exact symmetry of all known laws of physics, and, therefore, the qubit Hamiltonian $\hat{H}$, which corresponds to a real physical system, has to honor this symmetry as well. Therefore, the unitary operation describing evolution of the physical system $\hat{U}_{R}$ is generally known and represents a transformation which is inherited from the time-reversal symmetry of the original Hamiltonian $\mathcal{\hat H}$. In particular, if the qubit Hamiltonian $\hat{H}$ is real, then the corresponding evolution operator $\hat{U}(\tau) = e^{-i\hat{H}\tau/\hbar}$ is symmetric that entails $\hat{U}_R = \mathbf{1}$.

\begin{figure*}
\centering
\includegraphics[width=1.0\linewidth]{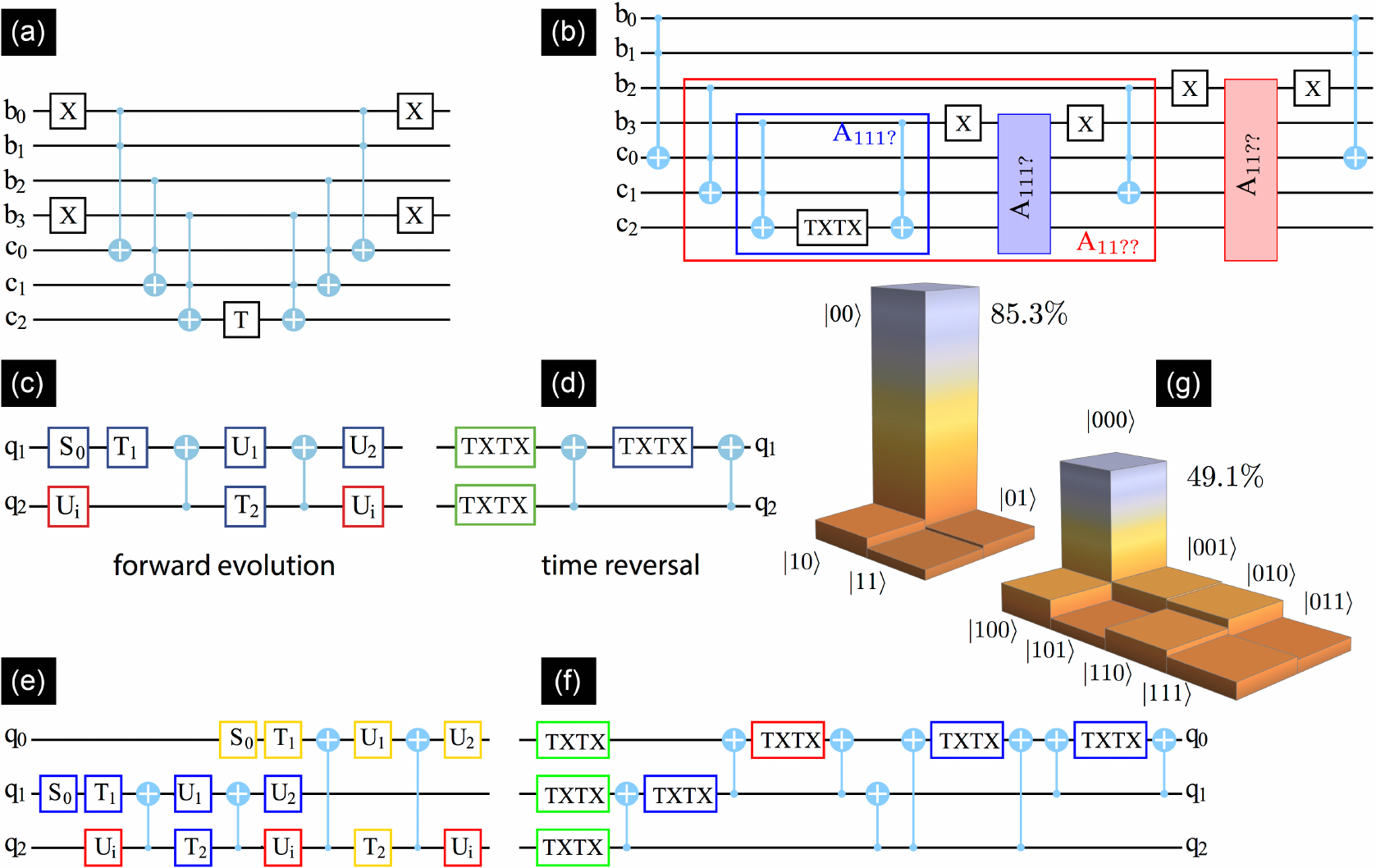}
\caption{Circuits realizing time reversal and the results of modeling. (a)~Quantum circuit that realizes the state-selective phase shift operation $\hat\Phi_{k=6}$ for a component $|0110\rangle$. The circuit involves three types of gates: $1$-qubit NOT gate $\hat{X}|b\rangle = |b\oplus 1\rangle$, $1$-qubit unitary rotation $\hat{T}(-2\varphi_k) [|0\rangle + a |1\rangle] = |0\rangle + a e^{-2i\varphi_k} |1\rangle$, and $3$-qubit Toffoli gate which reverts the state of the last target qubit if and only if two first control qubits are both set to $|1\rangle$: $\hat\Lambda_2|11\rangle \otimes|b\rangle = |11\rangle \otimes |b\oplus1\rangle$. The first three Toffoli gates set the ancillary qubit $c_2$ into $|1\rangle$ if and only if the qubit register is set to the $|0110\rangle$ state and the last three Toffoli gates restore the original state $|b_0b_1b_2b_3\rangle \otimes |000\rangle$. (b)~The quantum circuit with the optimal Toffoli gate arrangement which conjugates four components: $|1111\rangle$, $|1110\rangle$, $|1101\rangle$ and $|1100\rangle$. The circuit is partitioned into several nested blocks (subroutines) $A_{11??} \supset A_{111?}$, the question marks standing for an unknown bit value. The first-level block (blue) $A_{111?}$ conjugates only computational states where three senior qubits $|b_0b_1b_2\rangle$ are all set to $|1\rangle$. The next-level block (red) $A_{11??}$ contains as a subroutine the block $A_{111?}$ and conjugates all components $|b_0b_1\rangle = |11\rangle$. (c) and (e) The quantum circuits which model the scattering process of one or two particles (qubit lines $q_0$ and $q_1$) on the two level impurity (qubit line $q_2$). Unitary operations $\hat{U}_\mathrm{i}$ (red boxes) describe free evolution of the TLI during the time $\tau$. Remaining operations simulate the particle's scattering: the group of $1$-qubit gates (blue or yellow boxes) combined with the two CNOT gates implements the scattering operator $\hat{S}_\psi$ for the $q_1$ or $q_0$ particle. The parameters of the gates are adjusted in the specific way: $ \hat{T}_2 |1\rangle \otimes \hat{U}_2 \hat\sigma_x \hat{U}_1 \hat\sigma_x \hat{T}_1 |q\rangle = |1\rangle \otimes \hat{S}_2S_1^\dagger |q\rangle$ and $\hat{T}_2^\dagger |1\rangle \otimes \hat{U}_2 \hat{U}_1 \hat{T}_1|q\rangle = |1\rangle \otimes |q\rangle$, {\ref see Appendix~\ref{sec:phase_shifts}.} (d,f)~The $2$- and $3$-qubit quantum circuits realizing the exact complex conjugation procedure. A single qubit gate $\mathrm{TXTX}(\varphi,\bar\varphi) \equiv \hat{T}(\varphi) \hat\sigma_x \hat{T}(\bar\varphi) \hat\sigma_x $ performs a phase shift of a qubit components: $\mbox{TXTX}(\varphi,\bar\varphi) (a|0_i\rangle + b|1_i\rangle) = a e^{i\bar\varphi} |0_i\rangle + b e^{i\varphi} |1_i\rangle$, $i=0,1,3$. The gates $\mbox{TXTX}$ combined properly with the CNOT gates perform the controlled phase shifts associated with the single-qubit ($b_i$, $\bar{b}_i$, $i=0,1,2$) (green boxes), two-qubit ($b_i \oplus b_j$, $\overline{b_i \oplus b_j}$, $i,j =0,1,2$, $i<j$) (blue boxes) and three-qubit ($b_0\oplus b_1 \oplus b_2$, $\overline{b_0\oplus b_1 \oplus b_2}$) logical term (red box) in the Eq.~\eqref{eq:nAND}. (g)~Realization of the $2$ or $3$-qubit time reversal experiment performed on the IBM public quantum computer. The histogram shows (in percents) the appearance rates of the computational basis states obtained by the 8192 independent runs of the experiment.}
\label{fig:circuits}
\end{figure*}

In the following we assume the unitary $\hat{U}_R$ to be known and focus on the unitary implementation of a complex conjugation operation $\hat{K}$, $\hat{K} \to \hat{U}_\psi$. In particular, we quantify a complexity of such implementation as a number of elementary quantum gates or/and auxiliary qubits needed to implement $\hat{U}_\psi$. For a sparse coding scheme, the complex conjugation of the $\mathcal{N}$-qubit state $|\psi\rangle = \sum_{i=0}^{\mathcal{N}-1} |\psi_i| e^{i\varphi_i}\, |0_0\ldots 1_i \ldots 0_{\mathcal{N}-1}\rangle$ can be accomplished by the unitary operation $\hat{U}_{\psi}^{(1)} = \prod_{i=0}^{\mathcal{N}-1} \otimes\hat{T}_i(-2\varphi_i)$ where $\hat{T}_i(\varphi)$ is the single qubit operation: $\hat{T}_i(\varphi)|0_i\rangle = |0_i\rangle$ and $\hat{T}_i(\varphi)|1_i\rangle = e^{i\varphi}|1_i\rangle$. Consequently, the sparse coding scheme does not require the most ``expensive'' two-qubit gates at all but do require a large number $\mathcal{N}$ of qubits. For the dense coding scheme the situation is the opposite: this scheme involves only a logarithmically smaller number $n$ of qubits but instead requires implementation of $2^n$ $n$-qubit conditional phase shift operations: $\hat{K} \to \hat{U}_\psi^{(2)} = \sum_{j=0}^{2^n-1} |j\rangle \langle j| e^{-2i\varphi_j}$ which add proper phases to each component of the state $|\psi\rangle$: $\hat{U}_\psi^{(2)}|\psi\rangle = |\psi^*\rangle$. Therefore, $\hat{U}_\psi^{(2)}$ must involve two-qubit gates, i.e. conditional-NOT (CNOT) gates. We quantify the complexity of the dense coding scheme by a number $N_\oplus$ of CNOT gates needed to implement it. Each phase shift operation $\hat\Phi_i(\varphi) \equiv |i\rangle \langle i| e^{i\varphi}$ can be build with the help of $n-1$ ancillary qubits and $2(n-1)$ Toffoli gates, as shown in Fig.~\ref{fig:circuits}(a). In total, it requires $N_\oplus [\hat{U}_\psi^{(2)}] = 12(n-1)2^n \sim 12 N\log_2(N)$ CNOT gates. However, a proper arrangement of $\hat\Phi_i$ operations shown in Fig.~\ref{fig:circuits}(b) can reduce this number to be linear in $\mathcal{N}$: $N_\oplus \sim 24 \mathcal{N}$, see Appendix~\ref{sec:phase_shifts}. We thus arrive at the conclusion: \textit{The number of elementary operations needed for the exact time reversal procedure of the dynamics of a quantum system which on course its evolution sweeps a Hilbert space of a dimension $\mathcal{N}$ is bounded from above by some number $\mathcal{O}(\mathcal{N})$.}

\section{Time reversal experiment}

Now we are equipped to carry out an experiment implementing two- and three-qubit time-reversal procedures utilizing the public IBM quantum computer. We model a one dimensional particle scattering on a two-level impurity (TLI). The dynamics of the impurity is governed by a Hamiltonian, $\hat{H}_\mathrm{i} = \hbar\omega \bigl( \cos\alpha\, \hat\sigma_z + \sin\alpha\, \hat\sigma_x \bigr)$. The scattering potential seen by the particle depends on the state of the TLS. The corresponding scattering operator has the form $\hat{S}_\psi = |0\rangle \langle 0| \otimes \hat{S}_0 + |1\rangle \langle 1| \otimes \hat{S}_1$, where $\hat{S}_0$ and $\hat{S}_1$ are symmetric unitary scattering matrices of the TLI in a state $|0\rangle$ or $|1\rangle$.

This scattering problem is modeled by the evolution of the qubit register $\hat{U}_{n\mathrm{bit}}|q_\mathrm{i}\rangle \otimes \bigl(|q_1\rangle \otimes \cdots \otimes |q_n\rangle\bigr)$, where $|q_\mathrm{i}\rangle$ qubit describes the state of the TLI and the remaining qubits describe the state of scattered particles. The basis states $|0_i\rangle$ and $|1_i\rangle$, $i=1,\ldots n$ correspond to the left and right incoming/outgoing states of the $i$th particle. We consider the processes in which one or two incoming particles are scattered on the freely evolving TLI. The corresponding $2$-qubit and $3$-qubit evolution operators have the form: $\hat{U}_\mathrm{2bit} = \hat{U}_\mathrm{i}(\tau) \cdot \hat{S}_\psi^{(1)} \cdot \hat{U}_\mathrm{i}(\tau)$ and $\hat{U}_\mathrm{3bit} = \hat{U}_\mathrm{i}(\tau)\cdot \hat{S}_\psi^{(2)} \cdot \hat{U}_\mathrm{i}(\tau) \cdot \hat{S}_\psi^{(1)} \cdot \hat{U}_\mathrm{i}(\tau)$, where $\hat{U}_\mathrm{i}(\tau) = e^{-i\hat{H}_\mathrm{i}\tau/\hbar}$ describes the free evolution of the TLI, and $\hat{S}_\psi^{(i)}$, $i=1,2$ is the scattering operator for the $i$th particle. The corresponding quantum circuits realizing $\hat{U}_\mathrm{2bit}$ and $\hat{U}_\mathrm{3bit}$ are shown on Figs.~\ref{fig:circuits}(c) and \ref{fig:circuits}(e), see the details in Appendix~\ref{sec:time_reversal_algorithm}.

The $2$-qubit scattering model is endowed with the symmetric evolution operator $\hat{U}_\mathrm{2bit}$ and, therefore, its time reversal requires only the complex conjugation operation $\hat{R} = \hat{K}$. At variance, the evolution operator $\hat{U}_\mathrm{3bit}$ of the $3$-qubit model is non symmetric and its time reversal requires an additional unitary rotation $\hat{R} = \hat{U}_R \hat{K}$. It follows from the relation $\mathrm{SWAP}_{12} \cdot \hat{U}_\mathrm{3bit} \cdot \mathrm{SWAP}_{12} = \hat{U}_\mathrm{3bit}^t$, where $\mathrm{SWAP}_{12}|q_1\rangle \otimes |q_2\rangle = |q_2\rangle \otimes |q_1\rangle$ is the swap operation, that the required unitary operation $\hat{U}_R = \mathrm{SWAP}_{12}$.

According to the results of the Sec.~\ref{sec:reversal_algorithms}, the unitary implementation of the complex conjugation for a $2$- or $3$-qubit register will require $48$ or $144$ CNOT gates. These numbers are beyond of the present capability of the IBM public quantum computer due to the finite error rate 1.5--2.5\% of its CNOT gates. Here we utilize an alternative to Sec.~\ref{sec:reversal_algorithms} approach (see Appendix~\ref{sec:time_reversal_algorithm} for details), which is based on the arithmetic representation of the $n$-bit AND Boolean function,\cite{Barenco:1995}
\begin{multline}
	b_0 \wedge b_1 \wedge \ldots \wedge b_{n-1} 
	= \frac1{2^{n-1}} \biggl( \sum_{i_1} b_{i_1} 
	- \sum_{i_1<i_2} b_{i_1} \oplus b_{i_2}
	\\
	+ \sum_{i_1 < i_2 < i_3} b_{i_1} \oplus b_{i_2} \oplus b_{i_3} + \ldots 
	\\
	+ (-1)^{n-1} b_0 \oplus \ldots \oplus b_{n-1} \biggr).
	\label{eq:nAND} 
\end{multline}
This approach is more efficient at small $n$ since it does not need an ancillary qubits at all and requires $(n-1)2^{n-1}$ CNOT gates for the complex conjugation of an arbitrary $n$-qubit state that wins over the approach discussed in Sec.~\ref{sec:reversal_algorithms} for $n\leq 48$. In particular, at $n=2$ and $3$ one needs only two or eight CNOT gates, respectively. The corresponding $2$- and $3$-qubit quantum circuits are shown on Figs.~\ref{fig:circuits}(c) and \ref{fig:circuits}(f).

The time-reversal experiment runs in several steps: (i) the qubit register that is initially set into the state $|\psi(0)\rangle = |0\ldots 0\rangle$ accomplishes the forward time unitary evolution $|\psi_0\rangle \to |\psi_1\rangle = \hat{U}_{n\mathrm{bit}}|\psi_0\rangle$. Next, (ii$^\prime$) the unitary complex conjugation operation $\hat{K} = \hat{U}_\psi$ is applied $|\psi_1\rangle \to |\psi_1^*\rangle = \hat{U}_\psi |\psi_1\rangle$ followed by (ii$^{\prime\prime}$) the unitary transformation $\hat{U}_R$, $|\psi_1^*\rangle \to |\hat{R}\psi_1\rangle = \hat{U}_R |\psi_1^*\rangle$. As a result, the time-reversed state $|\hat{R}\psi_1\rangle$ is generated. Finally, at step (iii) one applies the same forward time unitary evolution $|\hat{R}\psi_1\rangle \to |\tilde\psi_0\rangle = \hat{U}_{n\mathrm{bit}}|\hat{R}\psi_1\rangle$ and measures the resulting state of the register in the computational basis. In practice, the step 2$^{\prime\prime}$ is only needed for the $3$-qubit model where $\hat{U}_R = \mathrm{SWAP}_{12}$ requires three additional CNOT gates. In order to save this number of CNOTs we replace the forward evolution operator $\hat{U}_\mathrm{3bit}$ at step (iii) by the new evolution operation obtained from $\hat{U}_\mathrm{3bit}$ via the physical interchange of two particle qubits, rather than to implement the $\mathrm{SWAP}_{12}$ operation at step (ii$^{\prime\prime}$).

At the end of the experiment, the above time reversal experiment sets the qubit register again into the initial state $|0\ldots 0\rangle$ with the probability unity, provided all quantum gates are prefect and no decoherence and relaxation processes are present. The exemplary outcome probabilities $P_{ij} = |\langle b_ib_j|\tilde\psi_0\rangle|^2$ and $P_{ijk} = |\langle b_ib_jb_k|\tilde\psi_0\rangle|^2$, $i,j,k =0,1$ obtained in a real experiment for the $2$- and $3$-qubit models are shown on the Fig.~\ref{fig:circuits}(g). One can see that the probability for observing the correct final state $|0\ldots0\rangle$ is less than $100\%$ and for $2$- and $3$-qubit experiment are given by $85.3\pm 0.4\%$ and $49.1 \pm 0.6\%$ correspondingly. This considerable distinction from the perfect scenario comes from the three main sources: (i)~the finite coherence time $T_2$ of qubits, (ii)~the errors of CNOT gates, and (iii)~the readout errors of the final state of the qubit register.

The observed outcome probabilities were obtained after $8192$ runs of each experiment at the same state of the `ibmqx4' $5$-qubit quantum processor, see details in Appendix~\ref{sec:reversal_experiment}. For the $2$-qubit experiment two processor's qubit lines $q_1$ and $q_2$ with the coherence times $41.0\mu$s and $43.5\mu$s and readout errors $\epsilon_{r1}=3.3\%$ and $\epsilon_{r2}= 2.9\%$ were involved. For the $3$-qubit experiment, the additional $q_0$ qubit line with $T_2 = 39.4\mu$s and the readout error $\epsilon_{r0} = 4.8\%$ was used. The $2$-qubit experiment requires six $\mathrm{CNOT}_{q2,q1}$ gates with the gate error $\epsilon_{g21} = 2.786\%$, while the $3$-qubit experiment acquires, in addition, six $\mathrm{CNOT}_{q2,q0}$ and four $\mathrm{CNOT}_{q1,q0}$ gates with the corresponding gate errors $\epsilon_{g20} = 2.460\%$ and $\epsilon_{g10} = 1.683\%$. This numbers give us a rough estimate of the net error rate for each experiment: $\epsilon_\mathrm{2bit} = 1 - (1-\epsilon_{g21})^6 (1-\epsilon_{r1})(1-\epsilon_{r2}) \approx 15.6\%$ and $\epsilon_\mathrm{3bit} = 1 - (1-\epsilon_{g21})^6 (1-\epsilon_{g20})^6 (1-\epsilon_{g10})^4 (1-\epsilon_{r0})(1-\epsilon_{r1})(1-\epsilon_{r2}) \approx 34.4\%$. One can see, that while this estimate agrees with an observed error of a $2$-qubit experiment, the error probability for the $3$-qubit experiment is underestimated. We argue that a time duration of a single $3$-qubit experiment is about $7.5\mu$s is comparable with $T_2$ times, while a single $2$-qubit experiment takes less time about $3\mu$s. Hence, the decoherence effects are more prominent in a $3$-qubit experiment that might explain the underestimated value of the $3$-qubit time-reversal experiment. The more experimental data for the different system parameters and processor states are discussed in Appendix~\ref{sec:reversal_experiment}.

\section{Conclusions}

Our findings break ground for investigations of the time reversal and the backward time flow in real quantum systems. One of the challenging directions to pursue, is the time dependence of the reversal complexity $\mathcal{N}$ of an evolving quantum state. In our work, we have shown that an isolated $d$-dimensional quantum particle with quadratic spectrum exhibits a polynomial complexity growth $\mathcal{N}(\tau) = \tau^d$. Uncovering the $\mathcal{N}(\tau)$ dependence for realistic situations, accounting for the interactions will establish a mechanism and the corresponding time-scale on which time-reversed states can spontaneously emerge. Another fundamental question is whether it is possible at all to design a quantum algorithm that would perform time-reversal more efficiently than using $\mathcal{O}(\mathcal{N})$ elementary gates? So far, our time-reversal schemes were scrolling one by one through the state components but did not exploit a quantum parallelism in its full power.

\paragraph*{Acknowledgements.}  This work was supported by U.S. Department of Energy, Office of Science, Materials Sciences and Engineering Division (G.B.L., A.V.L., V.M.V.); RFBR Grants 17-02-00396A (G.B.L.) and 18-02-00642A (A.V.L, G.B.L.); Pauli Center for Theoretical Studies at ETH Zurich (G.B.L.); Foundation for the Advancement of Theoretical Physics BASIS (G.B.L.); Swiss National Foundation through the NCCR QSIT (A.V.L.); Ministry of Education and Science of the Russian Federation 16.7162.2017/8.9 (A.V.L.); and Government of the Russian Federation through Agreement 05.Y09.21.0018 (G.B.L., A.V.L.). We gratefully acknowledge the IBM-Q team for providing us with access to their 5-qubit platform.

\appendix
\onecolumngrid

\section{Wave-packet reversal complexity} \label{sec:wave_packet_reversal}

Let a charged particle have one dimensional wave function $\psi(x) \equiv \sqrt{p(x)} e^{i\varphi(x)}$. Consider a fluctuating electromagnetic field potential $V(x,t)$ of the electromagnetic field which is approximated by the $N$-cell stepwise function $V(x,t) = \sum_{n=1}^N I_n(x) V(x_n,t)$, where $I_n(x)$ is an indicator function of the cell with the index $n$. Let us assume that during the short time interval a relatively strong non-homogenous fluctuation has emerged and the wave packet $\psi(x)$ acquires the coordinate dependent phase shift $\psi(x) \to \tilde\psi(x) = \psi(x) \exp(i\sum_n I_n(x) \phi_n)$, where $\phi_n = \int dt\, eV(x_n,t)/\hbar$. Consider then the specific fluctuation with $\phi_n(x) = -2\varphi(x_n)$ which drives the original wave packet $\psi(x)$ into its approximate complex conjugated form $\tilde\psi^*(x)$. The accuracy of such a conjugation procedure is defined through the overlap of the exact conjugated state $\psi^*(x)$ with the approximate conjugated state $\tilde\psi^*(x)$, $S = \langle \psi^*(x)| \tilde\psi^*(x)\rangle$,
\begin{equation}
	S = \sum_{n=1}^\infty \int dx\, I_n(x) p(x) \, e^{2i(\varphi(x) - \varphi(x_n))}.
\end{equation}
Then the probability of the correct reversion is given by $|S|^2$. Assuming that the particle density $p(x)$ changes slowly on the scale of large fluctuations of the particle phase $\varphi(x)$ one arrives at
\begin{equation}
	|S|^2 \approx 1 - \frac13 \sum_{n=1}^N p(x_n) \delta x_n\, \bigl[ \varphi^\prime (x_n) \delta x_n \bigr]^2,
	\label{eq:overlap}
\end{equation}
for the sufficiently small $\delta x_n$ of the cells defined through the condition $g(x_n) \equiv \varphi^\prime(x_n) \delta x_n \ll 1$. Then the error probability $\epsilon$ of the incorrect conjugation of the wavepacket is given by $|S|^2 = 1 - \epsilon$ and in the continuous limit one has
\begin{equation}
	\epsilon = \frac13 \int dx\, p(x)\, g^2(x).
	\label{eq:error}
\end{equation}
Let us now find the number of cells $N$ needed to approximate the electromagnetic field complex conjugation procedure with a given error probability level $\epsilon$. From the definition $g(x) = \varphi^\prime(x) \delta x$ one has
\begin{equation}
	N = \int dx\, \frac{|\varphi^\prime(x)|}{g(x)}.
\end{equation}
Minimizing the functional $N[g(x)]$ under the constraint Eq.~\eqref{eq:error} one finds
\begin{equation}
	N = \Bigl( \frac{\lambda^3(\psi)}{3\epsilon} \Bigr)^{1/2}, \quad \lambda(\psi) \equiv \int dx\, \bigl( |\psi(x)|^2 [\phi^\prime(x)]^2 \bigr)^{1/3}.
\end{equation}
Generalization of the above result to a $d$-dimensional case is straightforward
\begin{equation}
	N_d 
	= \lambda(\psi) \Bigl( \frac{\lambda(\psi)}{3\epsilon} \Bigr)^{d/2}, \quad 
	\lambda(\psi) 
	= \int d^d \vec{x}\, \bigl( |\psi(\vec{x})|^2 \vec\nabla^2 \varphi(\vec{x})\, \bigr)^{d/(d+2)}.
\end{equation}
Applying these results to the wave packet given by the Eq.~\eqref{eq:wave_packet}, one obtains $\lambda(\Psi) \sim (\hbar\tau/m)^{2/3} \int dx (f^2(x) x^2)^{1/3}$. We assume that initially at $\tau = 0$ the wave packet has the size $L_0$, so that $f^2(k) \sim L_0$ for $|k| \leq 1/L_0$ and, therefore, $\lambda(\Psi) \sim (L_\tau/L_0)^{2/3}$ where $L_\tau = \hbar\tau/mL_0$ is the size of the wave packet after the free evolution during the time $\tau$. Therefore, the number of the elementary cells needed to arrange the electromagnetic potential fluctuation which reverses the dynamics of a one dimensional wave packet is linear in $\tau$ since $N\sim \epsilon^{-1/2}\,L_\tau/L_0$, see also Ref.~\onlinecite{Lesovik:2013}. For a $d$-dimensional wave packet the number of cells grows polynomially with $\tau$ as $N \sim \epsilon^{-d/2}\,(L_\tau/L_0)^d$.

\section{Reversal of the qubit register dynamics} \label{sec:qubit_reversal}

Let the forward time dynamics of the $n$-qubit register state $|\psi(t)\rangle = \sum_{i=0}^{N-1} \psi_i(t) |i\rangle$ be governed by the Hamiltonian $\hat{H}$, $i\hbar \partial_t |\psi(t)\rangle = \hat{H}|\psi(t)\rangle$. The time-reversal symmetry of the Schr\"{o}dinger equation implies that if there is a forward time solution $|\psi(t)\rangle$ then the backward time solution $|\tilde\psi(t)\rangle$
\begin{equation}
	-i\hbar|\tilde\psi(t)\rangle = \hat{H} |\tilde\psi(t)\rangle
	\label{eq:schroedinger}
\end{equation}
also exists and is uniquely defined through the forward time solution via the time-reversal operation $\hat{R}$ such that $|\tilde\psi(t)\rangle = \hat{R} |\psi(t)\rangle$. The time-reversal operation $\hat{R}$ is an anti-unitary operation: $\langle \hat{R}\psi_1| \hat{R}\psi_2\rangle = \langle \psi_1|\psi_2\rangle^*$ and can be presented as a product $\hat{R} = \hat{U}_R \hat{K}$ of some unitary operator $\hat{U}_R$ and the complex conjugation operation $\hat{K}$ which we define with respect to the computational basis $|i\rangle$ of the qubit register as
\begin{equation}
	\hat{K}\Bigl( \sum_i \psi_i |i\rangle \Bigr) 
	= \sum_i \psi_i^*|i\rangle.
\end{equation}
Substituting $|\tilde\psi(t)\rangle = \hat{U}_R\hat{K}|\psi(t)\rangle$ into Eq.\,\eqref{eq:schroedinger} one finds
\begin{equation}
	i\hbar \partial_t |\psi(t)\rangle 
	= \bigl( \hat{U}_R^\dagger \hat{H} \hat{U}_R\bigr)^* |\psi(t)\rangle,
\end{equation}
and therefore the unitary operation $\hat{U}_R$ has to satisfy a relation,
\begin{equation}
	\hat{H} = \bigl( \hat{U}_R^\dagger \hat{H} \hat{U}_R \bigr)^*.
	\label{eq:UR}
\end{equation}
The relation~\eqref{eq:UR} defines the unitary $\hat{U}_R$. Indeed, the hermitian operator $\hat{H}$ can be represented in a form $\hat{H} = \hat{U}_H^\dagger \hat{E} \hat{U}_H$, where $\hat{E}$ is a real diagonal operator and $\hat{U}_H$ is unitary. Then it follows from the Eq.~\eqref{eq:UR}
\begin{equation}
	\hat{U}_R = \hat{U}_H^\dagger \hat{U}_H^*.
	\label{eq:UR2}
\end{equation}
The forward time evolution operator $\hat{U}(\tau) = e^{-i\hat{H}\tau/\hbar}$ applied to the time reversed state $|\tilde\psi(\tau)\rangle$ drives it into the new state
\begin{equation}
	\hat{U}(\tau)|\tilde\psi(\tau)\rangle = \hat{R} |\psi(0)\rangle.
	\label{eq:reversal1}
\end{equation}
Indeed,
\begin{align}
	\hat{U}(\tau)|\tilde\psi(\tau)\rangle 
	& \equiv e^{-i \hat{H}\tau / \hbar} \, \hat{U}_R \, \hat{K} \, e^{- i \hat{H}\tau / \hbar} \, |\psi(0)\rangle
	\nonumber
	\\
	& = e^{- i \hat{H}\tau / \hbar} \, e^{i \hat{U}_R \hat{H}^* \hat{U}_R^\dagger \tau / \hbar} \, \hat{R} \, |\psi(0)\rangle.
\end{align}
Making use of the explicit form of the $\hat{U}_R$ operator, see Eq.~\eqref{eq:UR2}, one has $\hat{U}_R \hat{H}^t \hat{U}_R^\dagger = \hat{H}$ that proves Eq.~\eqref{eq:reversal1}. Therefore, in order to restore the original state $|\psi(0)\rangle$ from the time-evolved state $|\psi(\tau)\rangle$ one has to apply the following sequence of operations
\begin{equation}
	|\psi(0)\rangle = \hat{R}^{-1} \hat{U}(\tau) \hat{R}\, |\psi(\tau)\rangle.
\end{equation}

\section{Optimal phase shifts arrangement} \label{sec:phase_shifts}

Here we outline an optimal arrangement of the state selective phase shift operations $\hat\Phi_i(\varphi) = |i\rangle \langle i| e^{i\varphi}$ entering the complex conjugation operation $\hat{U}_\psi = \prod_{i=0}^{2^n-1} \hat\Phi_i(-2\varphi_i)$ for the qubit state $|\psi\rangle = \sum_{i=0}^{2^n-1} |\psi_i| e^{i\varphi_i} |i\rangle$. Let us consider $2^{k-2}$ operations $\hat\Phi_k$ with index $k$ having the same values of two highest bits $b_0=b_1=1$: $k(k^\prime) = 2^{n-1}+2^{n-2}+k^\prime$, $k^\prime = 0,\ldots,2^{n-2}-1$. Then in the product $\prod_{k^\prime = 0}^{2^{n-2}-1} \hat\Phi_k(-2\varphi_k)$ one needs to check the values of the bits $b_0$ and $b_1$ only once, and this reduces the number of Toffoli gates. This recipe can be recursively repeated for the next lower bits $b_2$, $b_3$ and so on, see Fig.~\ref{fig:circuits}(b). Then the resulting quantum circuit comprises the sequence of nested blocks or subroutines $\mathcal{A}_{11b_2\ldots b_{n-1}} \supset \mathcal{A}_{111b_3\ldots b_{n-1}} \supset \ldots \supset \mathcal{A}_{1\ldots 1b_{n-1}}$ where each subroutine $\mathcal{A}_{1\ldots 1 b_{m} \ldots b_{n-1}}$ performs the controlled phase shift on all components $|k\rangle$ with first $m$ highest bits equal to $1$. As follows from the Fig.~\ref{fig:circuits}(b), the subroutine $\mathcal{A}_{11\ldots b_{m}\ldots b_{n-1}}$ involves two subroutines of the next lower level $\mathcal{A}_{11\ldots 1b_{m+1}\ldots b_{n-1}}$ and $\mathcal{A}_{11\ldots 0b_{m+1}\ldots b_{n-1}}$, and two additional Toffoli gates that are needed to check the value of the bit $b_{m+1}$. Therefore, the number of Toffoli gates $N_{\Lambda_2}[ \mathcal{A}_{11\ldots 1b_{m}\ldots b_{n-1}}]$ needed for the implementation of the subroutine $\mathcal{A}_{11\ldots 1b_{m}\ldots b_{n-1}}$ obeys the relation $N_{\Lambda_2}[\mathcal{A}_{11\ldots 1b_m\ldots b_{n-1}}] = 2 + 2N_{\Lambda_2} [ \mathcal{A}_{11\ldots 1b_{m+1}\ldots b_{n-1}}]$ with the boundary condition $N_{\Lambda_2} [ \mathcal{A}_{11\ldots 1b_{n-1}} ] = 2$, that gives $N_{\Lambda_2}[ \mathcal{A}_{11b_2\ldots b_{n-1}}] = 2^n-2$. The full $n$-qubit complex conjugation procedure $\hat{U}_\psi^{(2)}$ involves four different qubit subroutines $\mathcal{A}_{00b_2\ldots b_{n-1}}$, $\mathcal{A}_{01b_2\ldots b_{n-1}}$ and so on. This, finally, yields $N_{\Lambda_2}[ \hat{U}_\psi^{(2)}] = 4(2^n-2)$ and hence $N_\oplus[ \hat{U}_\psi^{(2)}] = 24(2^n-2) \sim 24\mathcal{N}$.

\section{Boolean function time-reversal algorithm} \label{sec:time_reversal_algorithm}

Here we describe the time-reversal procedure of a qubit register based on the arithmetic representation of a $n$-qubit Boolean function,
\begin{equation}
	b_{n-1} \wedge b_{n-2} \wedge \ldots \wedge b_0 
	= \left\{ \begin{array}{ll} 
		1,& b_0 = b_1 = \ldots = b_{n-1} = 1,\\
		0, & \mathrm{overwise.} 
	\end{array} \right.
\end{equation}
We find the minimal number of CNOT gates needed for the implementation of this procedure. Let us start with the two-qubit situation where one wishes to reverse the general two-qubit state $|\psi_2\rangle = e^{i\varphi_{00}}|00\rangle + e^{i\varphi_{01}}|01\rangle + e^{i\varphi_{10}}|10\rangle + e^{i\varphi_{11}}|11\rangle$. This requires to implement the complex conjugation procedure, which for a given state can be realized by the two-qubit unitary operation
\begin{equation}
	\hat{K}_2 
	= \sum_{b_0,b_1 = 0,1} e^{-2i\varphi_{b_1b_0}}\, |b_1b_0\rangle \langle b_1b_0| 
	\equiv e^{-2i\hat{F}(b_1,b_0)},
\end{equation}
where $\hat{F}(b_1,b_0)$ is the two-qubit Boolean function of the form
\begin{equation}
	\hat{F}(b_1,b_0) 
	= \varphi_{00} \,\bar{b}_1\wedge \bar{b}_0 + \varphi_{10}\, b_1 \wedge \bar{b}_0 
	+ \varphi_{01}\, \bar{b}_1 \wedge b_0 + \varphi_{11}\, b_1 \wedge b_0,
\end{equation}
and $\bar{b}_i$ denotes the logical negation of the bit $b_i$, $\bar{b}_i = \mathrm{NOT}(b_i)$. Making use of the arithmetic representation of $b_1 \wedge b_0$, see Eq.~\eqref{eq:nAND}, one finds
\begin{equation}
	\hat{F}(b_1,b_0) 
	= \frac{\varphi_{01} + \varphi_{11}}2 b_0 
	+ \frac{\varphi_{10} + \varphi_{00}}2 \bar{b}_0 
	+ \frac{\varphi_{10} + \varphi_{11}}2 b_1 
	+ \frac{\varphi_{01} + \varphi_{10}}2 \bar{b}_1
	- \frac{\varphi_{00} + \varphi_{11}}2 \, b_1 \oplus b_0 
	- \frac{\varphi_{10} + \varphi_{01}}2 \, \overline{b_1 \oplus b_0},
	\label{eq:F2}
\end{equation}
where $b_1 \oplus b_0$ denotes a bit summation by modulo 2,
\begin{equation}
	b_1 \oplus b_0 = \left\{ \begin{array}{ll} 0, & b_0 = b_1, \\
	1,& b_0 \neq b_1. \end{array} \right.
\end{equation}
The first four terms in the Eq.~\eqref{eq:F2} correspond to the one-qubit state dependent phase shifts and can be realized only via the single-qubit gates
\begin{equation}
	\hat{T}(\alpha) = \left( \begin{array}{cc}1&0\\0&e^{i\alpha} \end{array} \right), \quad
	\hat{X} = \left( \begin{array}{cc} 0&1\\1&0 \end{array} \right),
\end{equation}
available on the public IBM quantum computer. The last two-qubit terms in Eq.~\eqref{eq:F2} will require two-qubit CNOT gates. The overall quantum circuit which realizes the unitary operation $e^{-2i\hat{F}(b_1,b_0)}$ is described by the following sequence of unitary operations
\begin{multline}
	e^{-2i\hat{F}(b_1,b_0)}
	= \mbox{CNOT}_{0,1} 
	\cdot \bigl[\mbox{TXTX}_1( \varphi_{00} + \varphi_{11},\varphi_{10} +\varphi_{01}) \otimes \mathbf{1}_0\bigr] 
	\cdot \mbox{CNOT}_{0,1}
	\\
	\cdot \bigl[\mbox{TXTX}_1( - \varphi_{10} - \varphi_{11}, - \varphi_{00} - \varphi_{01}) 
	\otimes \mbox{TXTX}_0( - \varphi_{01} - \varphi_{11}, - \varphi_{10} - \varphi_{00}) \bigr],
\end{multline}
where $\mathrm{TXTX}_i(\varphi,\bar\varphi) \equiv \hat{T}_i(\varphi)\hat{X}_i \hat{T}_i(\bar\varphi) \hat{X}_i$ is a single-qubit unitary operation which adds specified phase shifts to the state components of the $i$th qubit: $\mathrm{TXTX}_i(\varphi, \bar\varphi) ( a|0_i\rangle + b|1_i\rangle ) = ae^{i\bar\varphi}|0_i\rangle + be^{i\varphi}|1_i\rangle$. The corresponding quantum circuit is shown in the Fig.~\ref{fig:circuits}(d) and involves only two $\mathrm{CNOT}_{0,1}$ gates, where $|b_0\rangle$ qubit serves as control bit and $|b_1\rangle$ as a target.

\begin{figure}
\centering
\includegraphics[width=0.3\linewidth]{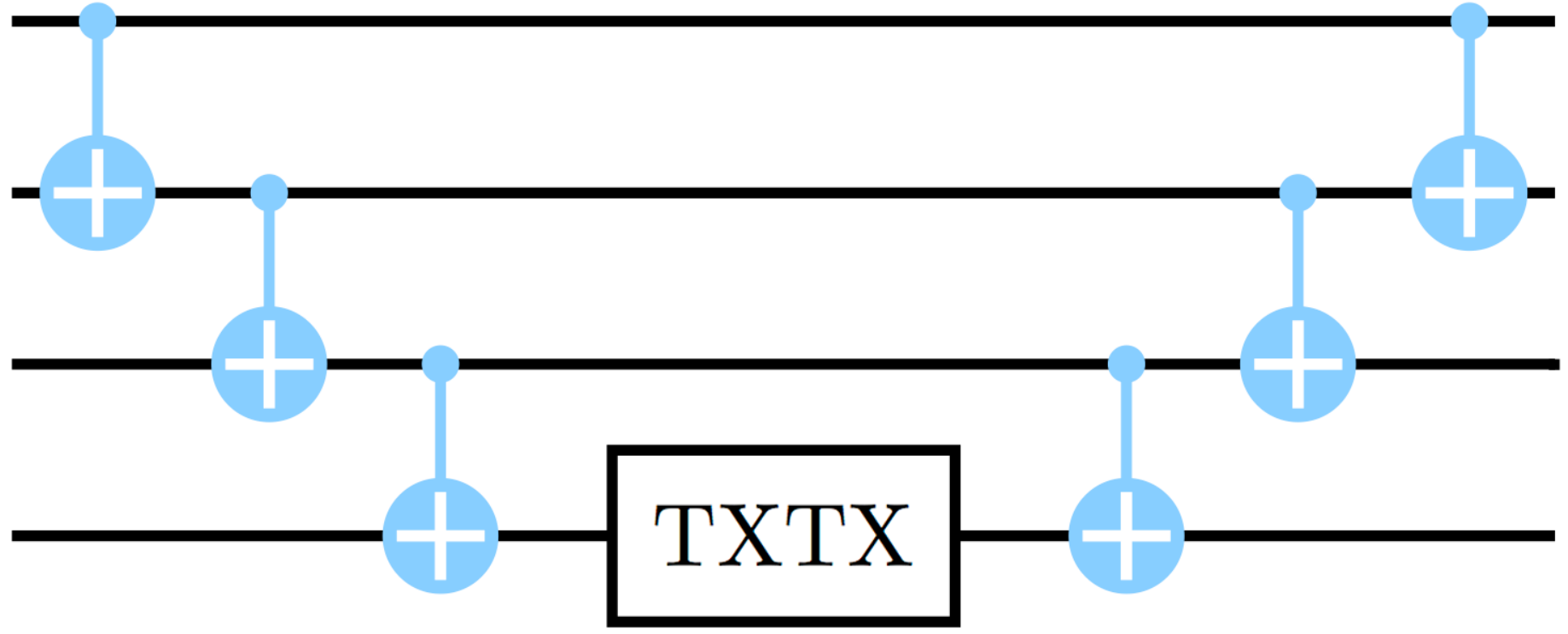}
\vspace{-2mm}
\caption{The quantum circuit which implements the four-qubit quantum gate $\mbox{CTXTX}_{0123}$. One can check that for any computational basis state $|b_3b_2b_1b_0\rangle$ the state of the elder bit $b_3$ is given by $b_0\oplus b_1 \oplus b_2 \oplus b_3$ right after the first ladder CNOT gates. The remaining symmetric half of CNOT gates is required in order to restore the original quantum state of the qubit register.}
\label{fig:nxor}
\end{figure}

The above two-qubit complex conjugation procedure can be further extended onto a general $n$-qubit state. As follows from the Eq.~\eqref{eq:nAND}, the quantum circuit performing complex conjugation of a given $n$-qubit state requires $n\choose{2}$ two-qubit operations,
\begin{equation}
	\mbox{CTXTX}_{i_1i_2}(\varphi,\bar\varphi) \equiv \left\{ \begin{array}{ll}
	\hat{T}_{i_2}(\varphi),& b_{i_1}\oplus b_{i_2} = 1\\
	\hat{T}_{i_2}(\bar\varphi),& b_{i_1}\oplus b_{i_2} = 0
	\end{array}
	\right., \quad 1\leq i_1 < i_2 \leq n,
\end{equation}
$n\choose{3}$ three-qubit operations,
\begin{equation}
	\mbox{CTXTX}_{i_1i_2i_3}(\varphi,\bar\varphi) \equiv \left\{ \begin{array}{ll}
	\hat{T}_{i_3}(\varphi),& b_{i_1}\oplus b_{i_2}\oplus b_{i_3} = 1\\
	\hat{T}_{i_3}(\bar\varphi),& b_{i_1}\oplus b_{i_2} \oplus b_{i_3} = 0
	\end{array}
	\right., \quad 1\leq i_1 < i_2 \leq i_3 \leq n,
\end{equation}
and so on. The general $n$-qubit operation $\mbox{CTXTX}_{i_1\ldots i_n}(\varphi,\bar\varphi)$, $1\leq i_1 < i_2 < \ldots < i_n$ is implemented with the help of $2(n-1)$ CNOT gates as shown on the Fig.~\ref{fig:nxor}. Therefore, one might conclude that in total $2\sum_{k=2}^n (k-1) {n\choose{k}} = 2^n(n-2)+2$ CNOT gates are required in order to implement a $n$-qubit time-reversal procedure.

However, the number of the CNOT gates can be reduced as far as some of operators $\mbox{CTXTX}_{i_1i_2\ldots}$ can be grouped together. Consider, for example, the unitary operation $\mbox{CTXTX}_{12}\cdot \mbox{CTXTX}_{123}$. Its straightforward implementation requires $4+2$ CNOT gates. A more savvy arrangement is shown in the Fig.~\ref{fig:circuits}(f). There the computational state of the second qubit $b_1$ right after the first $\mbox{CNOT}_{0,1}$ gate is given by $b_1\oplus b_2$. This enables one to implement the controlled phase shift $\mbox{CXTXT}_{01}$ right after the first $\mbox{CNOT}_{0,1}$ operation. At this moment, one need not to restore the original bit values $b_0$ and $b_1$ but rather to add the second $\mbox{CNOT}_{1,2}$, set the third qubit $b_3$ into the state $b_0\oplus b_1\oplus b_2$, and to implement the controlled phase shift $\mbox{CTXTX}_{012}$. Hence the unitary operation $\mbox{CTXTX}_{01}\cdot \mbox{CTXTX}_{012}$ will require the same number of CNOT gates as the operation $\mbox{CTXTX}_{012}$ alone. As a result, the complex conjugation operation of a given $3$-qubit state can be implemented using only 8 CNOT gates as shown in Fig.~\ref{fig:circuits}(f).

The above CNOT optimization technique can be easily generalized to a $n$-qubit case. Consider a product $\mbox{CTXTX}_{i_1i_2} \cdot \mbox{CTXTX}_{i_1i_2i_3} \cdot \mbox{CTXTX}_{i_1i_2i_3i_4} \ldots \cdot \mbox{CTXTX}_{i_1i_2i_3i_4\ldots i_n}$ where a sequence of nested strings of the qubit indices $i_1i_2 \subset i_1i_2i_3 \subset \ldots \subset i_1i_2i_3i_4\ldots i_n$ are formed by adding an additional index to the right hand side of a previous string. Then the implementation of this product requires the same number of CNOT gates as the largest $\mbox{CTXTX}_{i_1i_2i_3i_4\ldots i_n}$ factor of the product. This observation lets us find a number of CNOT gates $N_\oplus[\hat{K}_n]$ needed to implement the complex conjugation unitary operation $\hat{K}_n$ of a given $n$-qubit state.

Let us assume that $N_\oplus[ \hat{K}_{n-1} ]$ for a $n-1$ qubit register $b_1\ldots b_{n-1}$ is known. Let us add an additional qubit line $b_0$ and find how many additional operations $\mbox{CTXTX}(i_1i_2\ldots)$ one needs in order to complete the complex conjugation task for $n$-qubit register $b_0\ldots b_{n-1}$. Obviously any such additional operation $\mbox{CTXTX}_{s}$ has its parameter string $s = i_1 \ldots i_k$ starting from the index $0$, i.e. $i_1 = 0$. Consider for example $n=4$ case. Then there are seven additional operations,
\begin{equation}
	\mbox{CTXTX}_{0123} \cdot \mbox{CTXTX}_{012} \cdot \mbox{CTXT}_{013} \cdot \mbox{CTXTX}_{02} \cdot \mbox{CTXTX}_{01} \cdot \mbox{CTXTX}_{02} \cdot \mbox{CTXTX}_{03}.
\end{equation}
Making an optimization procedure one can group these operations as
\begin{equation}
	\bigl( \mbox{CTXTX}_{01} \cdot \mbox{CTXTX}_{012} \cdot \mbox{CTXTX}_{0123} \bigr) 
	\cdot \bigl( \mbox{CTXTX}_{02} \cdot \mbox{CTXTX}_{023} \bigr) 
	\cdot \mbox{CTXTX}_{013} \cdot \mbox{CTXTX}_{03},
\end{equation}
and hence
\begin{equation}
	N_\oplus\bigl[ \hat{K}_4] 
	= N_{\oplus}\bigl[\mbox{CTXTX}_{0123}\bigr] + N_\oplus\bigl[ \mbox{CTXTX}_{023} \bigr] + N_\oplus\bigl[ \mbox{CTXTX}_{013} \bigr]
	+ N_\oplus\bigl[ \mbox{CTXTX}_{03} \bigr] + N_\oplus\bigl[ \hat{K}_3 \bigr],
\end{equation}
where $N_\oplus[\mbox{CTXTX}_s]$ is the number of CNOT gates needed for the operation $\mbox{CTXTX}_s$. One can note, that only generalized operations $\mbox{CTXTX}_s$ with the inputs strings $s = i_1 \ldots i_k$ where first and last indices are equal to $0$ and $3$, respectively are counted for the total number of the CNOT gates. Therefore, for a general case, the following relation holds
\begin{align}
	N_\oplus\bigl[ \hat{K}_n \bigr] 
	& = N_\oplus\bigl[ \hat{K}_{n-1} \bigr] + N_\oplus\bigl[\mbox{CTXTX}_{1n}\bigr] + \!\!\! \sum_{1<k_1<n} \!\!\!\! N_\oplus\bigl[\mbox{CTXTX}_{1k_1n}\bigr]
	 + \!\!\!\!\! \sum_{1<k_1<k_2<n} \!\!\!\!\!\!\!\! N_\oplus\bigl[\mbox{CTXTX}_{1k_1k_2n}\bigr] + \ldots + N_\oplus\bigl[\mbox{CTXTX}_{1\ldots n}\bigr]
	\nonumber \\
	& = N_\oplus\bigl[ \hat{K}_{n-1} \bigr] + \sum_{k=0}^{n-2} 2(k+1) {n-2\choose{k}} 
	= N_\oplus\bigl[ \hat{K}_{n-1} \bigr] + n2^{n-2} ,
\end{align}
and, therefore,
\begin{equation}
	N_\oplus\bigl[ \hat{K}_n \bigr] = (n-1)2^{n-1}, \quad n>1.
\end{equation}

\section{Simulation of scattering on a two-level impurity} \label{sec:two_level_impurity}

Here we discuss a spinless particle which scatters on a two-level impurity (TLI). The free dynamics of the TLI is governed by a Hamiltonian
\begin{equation}
	\hat{H}_\mathrm{i} 
	= \hbar\omega \bigl( \cos\alpha \, \hat\sigma_z + \sin\alpha \, \hat\sigma_x \bigr).
\end{equation}
The scattering process is described by the $2\times2$ scattering matrix $\hat{S}_i$, $i=0,1$ whose form depends on the impurity state. The quantum state of the particle-impurity system can be described as the two-bit state $|\psi\rangle = \sum_{b_0,b_1=0,1} A_{b_1b_0}|b_1\rangle \otimes |b_0\rangle$ where the first qubit describes the TLI and the second one describes the propagation direction of an incoming/scattered particle. Let the system start in the state $|\psi(0)\rangle = |0\rangle \otimes |L\rangle$ with the particle coming from the left. Let after the time $\tau>0$ the particle be scattered on the TLI. The resulting state $|\psi(\tau)\rangle$ is generated by the sequence of unitary operations $|\psi(\tau)\rangle = \hat{S}_\psi \cdot [ \hat{U}_\mathrm{i}(\tau) \otimes \mathbf{1} ] |\psi(0)\rangle$, where the unitary operator $\hat{U}_\mathrm{i}(\tau) \equiv e^{-i\hat{H}_\mathrm{i}\tau/\hbar}$ describes the free evolution of TLI and
\begin{equation}
	\hat{S}_\psi 
	= |0\rangle \langle 0| \otimes \hat{S}_0 + |1\rangle \langle 1| \otimes \hat{S}_1
\end{equation}
describes the state dependent scattering process of the incoming particle. The unitary operator $\hat{U}_\mathrm{i}(\tau) = e^{-i \omega\tau (\cos\alpha \, \hat\sigma_x + \sin\alpha \, \hat\sigma_z)}$ is symmetric. In the absence of the magnetic field, the scattering operator $\hat{S}_\psi$ is symmetric as well. Let the state freely evolve after the scattering at the $t=\tau$ during the same time period $\tau$. Then the resulting state $|\psi(2\tau)\rangle = [\hat{U}_\mathrm{i}(\tau) \otimes \mathbf{1}] |\psi(\tau)\rangle$ can be generated from the initial state $|\psi(0)\rangle$ by the \textit{symmetric} $2$-qubit unitary operator
\begin{equation}
	\hat{U}_\mathrm{2bit} 
	= \bigl[\hat{U}_\mathrm{i}(\tau) \otimes \mathbf{1}\bigr] 
	\cdot \hat{S}_\psi \cdot \bigl[\hat{U}_\mathrm{i}(\tau) \otimes \mathbf{1}\bigr].
	\label{eq:Utwo}
\end{equation}
Therefore, as we have already discussed in Appendix~\ref{sec:qubit_reversal}, the time reversal procedure of the $2$-qubit state $|\psi(2\tau)\rangle$ requires only the unitary implementation of the complex conjugation operation $|\psi(2\tau)\rangle \to |\psi^*(2\tau)\rangle$.

Our goal is to implement the unitary operation $\hat{U}_\mathrm{2bit}$ via the set of quantum gates available on the IBM public quantum computer. The only available two-qubit gate is the $\mathrm{CNOT}_{b_c,b_t}$ gate, where $b_c$ the qubit serves as a control and $b_t$ qubit serves a target. Among the standard $1$-qubit gates we will need two available generalized $1$-qubit gates: the relative phase shift gate $\hat{T}(\alpha)$, introduced in the Appendix~\ref{sec:time_reversal_algorithm} and the full $1$-qubit unitary rotation
\begin{equation}
	\hat{U}_3(\theta,\alpha,\beta) \equiv \hat{T}(\alpha) \cdot \hat{R}(\theta) \cdot \hat{T}(\beta),
	\label{eq:U3}
\end{equation}
where
\begin{equation}
	\hat{R}(\theta) = \left( \begin{array}{cc} 
		\cos\cfrac\theta2 & -\sin\cfrac\theta2 \vspace{1mm} \\ 
		\sin\cfrac\theta2 & \cos\cfrac\theta2
	\end{array} \right).
\end{equation}
Any $2\times 2$ unitary matrix $\hat{U}$ can be represented in the form\,\eqref{eq:U3} up to some phase factor: $\hat{U} = e^{i\delta} \hat{U}_3(\theta,\alpha,\beta)$. In particular, any symmetric $2\times 2$ unitary matrix $\hat{U} = \hat{U}^t$ has the form $e^{i\delta} \hat{U}_3(\theta, \alpha, \alpha +\pi)$. Therefore, a given set of matrices $\hat{U}_\mathrm{i}(\tau)$, $\hat{S}_0$ and $\hat{S}_1$ entering into the definition of the model can be presented as
\begin{align}
	& \hat{U}_\mathrm{i}(\tau) = e^{i\delta} \hat{U}_3(\xi,\eta,\eta+\pi), \\
	& \hat{S}_i = e^{i\delta_i}\hat{U}_3(\theta_i,\varphi_i,\varphi_i+\pi), \quad i = 0,1.
\end{align}
The phase exponent $e^{i\phi}$ gives only a trivial common phase factor for the system state and will be omitted in what follows. Without any loss of generality we assume $\delta_0 = 0$ as well.

Next, let us construct the $2$-qubit operation $\hat{S}_\psi$ using as less CNOT gates as possible. It turns out that $\hat{S}_\psi$ can be constructed with the help of only two CNOT gates. Indeed,
\begin{equation}
	\hat{S}_\psi 
	= \bigl( |1\rangle \langle 1| \otimes \hat{S}_1 \hat{S}_0^\dagger + |0\rangle\langle 0| \otimes \mathbf{1} \bigr) 
	\cdot \bigl[ \mathbf{1}\otimes \hat{S}_0\bigr]
	\equiv \Lambda_{b_1,b_0}(\hat{S}_1\hat{S}_0^\dagger) \cdot \bigl[ \mathbf{1}\otimes \hat{S}_0\bigr],
	\label{eq:SviaW}
\end{equation}
where $\Lambda_{b_1,b_0}(\hat{W})$ is a controlled $\hat{W}$-gate,
\begin{equation}
	\Lambda_{b_c,b_t}(\hat{W})\bigl(|b_c\rangle \otimes |b_t\rangle\bigr)
	= \left\{ \begin{array}{ll}
		|b_c\rangle \otimes \hat{W}|b_t\rangle, & b_1 = 1, \\
		|b_c\rangle \otimes |b_t\rangle, & b_1 = 0.
	\end{array} \right.
\end{equation}
The unitary matrix $\hat{W} = \hat{S}_1 \hat{S}_0^\dagger \equiv e^{i\delta} \hat{U}_3(\theta,\alpha,\beta)$ can be represented as,
\begin{equation}
	\hat{W} 
	= e^{i\delta + i(\alpha+\beta)/2} \,
	\hat{T}(\alpha) \, \hat{R}\Bigl(\frac\theta2\Bigr) \, \hat\sigma_x \, \hat{R}\Bigl(-\frac\theta2\Bigr) \,
	\hat{T}\Bigl(-\frac{\alpha+\beta}2\Bigr) \, \hat\sigma_x \, \hat{T}\Bigl(\frac{\beta-\alpha}2\Bigr).
	\label{eq:W2}
\end{equation}
The advantage of the latter representation is that if one replaces in the Eq.~\eqref{eq:W2} two Pauli matrices $\hat\sigma_x$ by the identity operator, one gets a phase shift $e^{i\delta + i(\alpha+\beta)/2}$ only. Therefore,
\begin{equation}
	\Lambda_{b_1,b_0}(\hat{W}) 
	= \Bigl[ \hat{T}\Bigl(\delta +\frac{\alpha+\beta}2\Bigr) \otimes \hat{U}_3\Bigl(\frac\theta2,\alpha,0\Bigr)\Bigr]
	\cdot \mathrm{CNOT}_{b_1,b_0} 
	\cdot \Bigr[\mathbf{1}\otimes\hat{U}_3\Bigl(-\frac\theta2,0,-\frac{\alpha + \beta}2\Bigr) \Bigr] 
	\cdot \mathrm{CNOT}_{b_1,b_0} 
	\cdot \Bigl[\mathbf{1} \otimes \hat{T}\Bigl(\frac{\beta - \alpha}2\Bigr)\Bigr],
	\label{eq:LW}
\end{equation}
and the whole evolution operator, see Eq.~\eqref{eq:Utwo} can be presented as,
\begin{equation}
	\hat{U}_\mathrm{2bit} 
	= \bigl[ \hat{U}_3(\xi,\eta,\eta+\pi) \otimes \mathbf{1} \bigr] 
	\cdot \Lambda_{b_1,b_0}(\hat{S_1}\hat{S}_0^\dagger)
	\cdot \bigl[ \hat{U}_3(\xi,\eta,\eta+\pi) \otimes \hat{U}_3(\theta_0,\varphi_0,\varphi_0+\pi)\bigr].
\end{equation}
The corresponding $2$-qubit quantum circuit is shown on a Fig.~\ref{fig:circuits}(c).

Similarly, we consider a $3$-qubit model describing the scattering of two particles on a TLI. We assume that particles arrive to the TLI with the time separation $\tau$,
\begin{multline}
	\hat{U}_\mathrm{3bit} = \bigl[\hat{U}_\mathrm{i}(\tau)\otimes \mathbf{1} \otimes \mathbf{1}\bigr] \cdot \bigl[ |0\rangle \langle 0| \otimes \mathbf{1} \otimes \hat{S}_0 + |1\rangle \langle 1| \otimes \mathbf{1} \otimes \hat{S}_1 \bigr]
	\\
	\cdot \bigl[\hat{U}_\mathrm{i}(\tau)\otimes \mathbf{1} \otimes \mathbf{1}\bigr] \cdot \bigl[ |0\rangle \langle 0| \otimes \hat{S}_0 \otimes \mathbf{1} + |1\rangle \langle 1| \otimes \hat{S}_1 \otimes \mathbf{1} \bigr] \cdot \bigl[\hat{U}_\mathrm{i}(\tau)\otimes \mathbf{1} \otimes \mathbf{1}\bigr],
\end{multline}
where the first (eldest) bit describes the state of the TLI and the second and third qubits describe the scattering state of the first and second particles correspondingly. The quantum circuit which implements the evolution operator $\hat{U}_\mathrm{3bit}$ is shown in the Fig.~\ref{fig:circuits}(e).

\section{Time-reversal experiment} \label{sec:reversal_experiment}

In the simulation experiment we choose fixed scattering matrices of the two-level impurity (TLI),
\begin{equation}
	\hat{S}_0 = \left[ \begin{array}{cc}
		\cfrac12 & \cfrac{\sqrt{3}}{2} \vspace{1mm} \\ 
		\cfrac{\sqrt{3}}{2} & -\cfrac12
	\end{array}\right],
	\quad
	\hat{S}_1 = \left[ \begin{array}{cc}
		\cfrac{\sqrt{3}}{2} & \cfrac12e^{i\pi/3} \vspace{1mm} \\ 
		\cfrac12 e^{i\pi/3} & -\cfrac{\sqrt{3}}{2}e^{2\pi i/3}
	\end{array}\right],
\end{equation}
for the $|0\rangle$ and $|1\rangle$ impurity states correspondingly. Then
the state dependent scattering operator $\hat{S}_\psi$, see Eqs.~\eqref{eq:SviaW} and \eqref{eq:LW}, is given by the following sequence of quantum gates,
\begin{multline}
	\hat{S}_\psi \approx \bigl[ \mathbf{1} \otimes \hat{U}_3(0.723,-1.27) \bigr] \cdot \mathrm{CNOT}_{1,2} \cdot \bigl[ \hat{T}(1.047) \otimes \hat{U}_3(-0.723,0,-0.523) \bigr]
	\\
	\cdot \mathrm{CNOT}_{1,2}\bigl[ \mathbf{1} \otimes \hat{T}(1.761)\bigr] \cdot\Bigl[ \mathbf{1} \otimes \hat{U}_3\Bigl(\frac{2\pi}3,0,\pi\Bigr) \Bigr],
\end{multline}
where the first (control) qubit describes a state of TLI and the second (target) qubit describes a scattering state of the particle, $\hat{U}_3(\alpha,\varphi,\lambda)$ and $\hat{T}(\varphi)$ are generalized one-qubit gates available on the IBM quantum computer.

The free evolution operator $\hat{U}_\mathrm{i}(\tau) = e^{-i\hat{H}_\mathrm{i}\tau/\hbar}$ with $\hat{H}_\mathrm{i} = \hbar\omega(\cos\alpha \, \hat\sigma_z + \sin\alpha \, \hat\sigma_x)$ is parameterized by two parameters $\omega\tau$ and $\alpha$. The unitary operator $\hat{U}_\mathrm{i}(\tau)$ is symmetric and for a fixed values of $\omega\tau$ and $\alpha$ can be presented in the form,
\begin{equation}
	\hat{U}_\mathrm{i}(\tau) 
	= e^{i\delta} \hat{U}_3(\xi,\eta,\eta+\pi), 
	\quad \xi = \xi(\omega\tau,\alpha), 
	\quad \eta = \eta(\omega\tau,\alpha),
\end{equation}
where $e^{i\delta}$ some phase factor which changes only an overall phase of the qubit register; $\xi$ and $\eta$ are parameters which uniquely defined by $\omega\tau$ and $\alpha$. In the following we choose $\omega\tau = \pi/6$ and vary the parameter $\alpha$ among four values $\pi/6$, $\pi/4$, $\pi/3$ and $\pi/2$ with the corresponding gate parameters,
\begin{align*}
	&\xi\Bigl( \frac{\pi}6,\frac\pi6\Bigr) \approx 0.505, \qquad \eta \Bigl( \frac{\pi}6,\frac\pi6\Bigr) \approx -1.107,
	\\
	&\xi\Bigl( \frac{\pi}6,\frac\pi4\Bigr) \approx 0.723, \qquad \eta \Bigl( \frac{\pi}6,\frac\pi4\Bigr) \approx -1.183,
	\\
	&\xi\Bigl( \frac{\pi}6,\frac\pi3\Bigr) \approx 0.896, \qquad \eta \Bigl( \frac{\pi}6,\frac\pi3\Bigr) \approx -1.290,
	\\
	& \xi\Bigl( \frac{\pi}6,\frac\pi2\Bigr) \approx 1.047, \qquad \eta \Bigl( \frac{\pi}6,\frac\pi2\Bigr) \approx -\pi/2.
\end{align*}

\begin{table}[H]
\caption{The occurrence rates of the computational basis states for $2$-qubit experiments.}
\label{tab:2qubit_rates}
\centering
\begin{tabular}{|c|c|c|c|c|c|c|}
\hline
$\omega\tau$&$\alpha$&$|00\rangle$&$|10\rangle$&$|01\rangle$&$|11\rangle$&$F$
\rule{0pt}{0ex} \rule[-1.25ex]{0pt}{0pt} \\
\hline
$\pi/6$&$\pi/6$&$6949$&$437$&$562$&$244$&$84.8\pm0.4\%$
\rule{0pt}{2ex} \\
\hline
$\pi/6$&$\pi/4$&$6916$&$440$&$576$&$260$&$84.4\pm0.4\%$
\rule{0pt}{2ex} \\
\hline
$\pi/6$&$\pi/3$&$6983$&$370$&$560$&$279$&$85.2\pm0.4\%$
\rule{0pt}{2ex} \\
\hline
$\pi/6$&$\pi/2$&$6950$&$338$&$551$&$353$&$84.8\pm0.4\%$
\rule{0pt}{2ex} \\
\hline
\end{tabular}
\end{table}

\begin{table}[H]
\caption{The occurrence rates of the computational basis states for $3$-qubit experiments.}
\label{tab:3qubit_rates}
\centering
\begin{tabular}{|c|c|c|c|c|c|c|c|c|c|c|}
\hline
$\omega\tau$&$\alpha$&$|000\rangle$&$|001\rangle$&$|010\rangle$&$|011\rangle$&$|100\rangle$&$|101\rangle$&$|110\rangle$&$|111\rangle$&$F$
\rule{0pt}{0ex} \rule[-1.25ex]{0pt}{0pt} \\
\hline
$\pi/6$&$\pi/6$&$3909$&$1380$&$1069$&$487$&$482$&$309$&$332$&$224$&$47.7\pm0.5\%$
\rule{0pt}{2ex} \\
\hline
$\pi/6$&$\pi/4$&$3934$&$1157$&$981$&$380$&$618$&$360$&$407$&$355$&$48.0\pm0.5\%$
\rule{0pt}{2ex} \\
\hline
$\pi/6$&$\pi/3$&$3957$&$832$&$884$&$327$&$859$&$359$&$531$&$443$&$48.3\pm 0.5\%$
\rule{0pt}{2ex} \\
\hline
$\pi/6$&$\pi/2$&$3879$&$355$&$1050$&$425$&$964$&$418$&$630$&$471$&$47.3\pm0.5\%$
\rule{0pt}{2ex} \\
\hline
\end{tabular}
\end{table}

\begin{table}[H]
\caption{Relaxation times $T_1$, coherence times $T_2$, readout errors $\epsilon_r$ and one-qubit gate errors $\epsilon_1$ for each qubit line.}
\label{tab:ibmqx4}
\centering
\begin{tabular}{|c|c|c|c|c|}
\hline
$q_n$&$T_1(\mu\mathrm{s})$&$T_2(\mu\mathrm{s})$&$\epsilon_r(\%)$&$\epsilon_1(\%)$
\rule{0pt}{0ex} \rule[-1.25ex]{0pt}{0pt} \\
\hline
$q_0$&$52.4$&$47.3$&$4.2$&$0.077$
\rule{0pt}{2ex} \\
\hline
$q_1$&$58.0$&$40.6$&$3.6$&$0.103$
\rule{0pt}{2ex} \\
\hline
$q_2$&$46.9$&$47.4$&$2.8$&$0.137$
\rule{0pt}{2ex} \\
\hline
\end{tabular}
\end{table}

The occurrence rates of the computational basis states for $2$-qubit and $3$-qubit experiments are shown in Tables~\ref{tab:2qubit_rates} and \ref{tab:3qubit_rates} for the different input parameters of the model. The $2$-qubit experiment used $q_1$ and $q_2$ qubit lines of the `ibmqx4' five qubit quantum processor. The $3$-qubit experiment used in addition a $q_0$ qubit line. In both experiments the $q_2$ qubit line has modeled a state of TLI. The calibration state of the quantum computer was the same for all experiments. The qubit's relaxation times $T_1$, coherence times $T_2$, readout errors $\epsilon_r$ and one-qubit gate errors $\epsilon_1$ for each qubit line are shown in the Table~\ref{tab:ibmqx4}. The errors of the CNOT gates $\mathrm{CNOT}_{q2,q0}$, $\mathrm{CNOT}_{q2,q1}$ and $\mathrm{CNOT}_{q1,q0}$ used in the experiments are $\epsilon_{g20} = 1.91\%$, $\epsilon_{g21} = 2.68\%$ and $\epsilon_{g10}= 1.70\%$ respectively. These processor's state parameters allows us to estimate a theoretical value of a time-reversal fidelity $F = |\langle 0\ldots 0| \tilde\psi_0\rangle|^2$, where $|\tilde\psi_0\rangle$ is a final state of the qubit register. For the used gate arrangement one has,
\begin{align}
	& F_\mathrm{2bit}^\mathrm{theor} 
	= (1-\epsilon_{g21})^6 (1-\epsilon_{r1})(1-\epsilon_{r2}) 
	\approx 79.6\%,
	\\
	& F_\mathrm{3bit}^\mathrm{theor} 
	= (1-\epsilon_{g21})^6(1-\epsilon_{g20})^6 (1-\epsilon_{g10})^4 
	(1-\epsilon_{r0})(1-\epsilon_{r1})(1-\epsilon_{r2}) 
	\approx 63.4\%,
\end{align}
while the experimentally observed values of the time-reversal fidelity are shown in Tables~\ref{tab:2qubit_rates} and \ref{tab:3qubit_rates}.

\end{document}